\documentclass[twocolumn,prd,nofootinbib,showpacs,floatfix,preprintnumbers]{revtex4}
\usepackage{epsfig}
\usepackage[dvips]{color}

\def\be{\begin{equation}}
\def\ee{\end{equation}}
\def\ba{\begin{eqnarray}}
\def\ea{\end{eqnarray}}

\def\lsim{\raise0.3ex\hbox{$\;<$\kern-0.75em\raise-1.1ex\hbox{$\sim\;$}}}
\def\gsim{\raise0.3ex\hbox{$\;>$\kern-0.75em\raise-1.1ex\hbox{$\sim\;$}}}

\newcommand{\bw}{\begin{widetext}}
\newcommand{\ew}{\end{widetext}}
\begin{document}

\title{$\!$Astrophysical interpretation of the medium
scale clustering in the ultrahigh energy sky}
\author{A.~Cuoco$^{1}$, G.~Miele$^{2,3}$, P.~D.~Serpico$^{4}$}

\affiliation{$^1$Institut for Fysik og Astronomi, Aarhus
Universitet Ny Munkegade, Bygn. 1520 8000 Aarhus Denmark \\
$^2$Universit\`{a} ``Federico II", Dipartimento di Scienze
Fisiche, Napoli, Italy \& INFN Sezione di Napoli\\
$^3$Instituto de F\'{\i}sica Corpuscular (CSIC-Universitat
de Val\`{e}ncia), Ed.\ Institutos de Investigaci\'{o}n, Apdo.\ 22085,
E-46071 Val\`{e}ncia, Spain.\\
$^4$ Center for Particle Astrophysics, Fermi National Accelerator
Laboratory, Batavia, IL 60510-0500 USA}
\date{\today}

\begin{abstract}
We compare the clustering properties of the combined dataset of
ultra-high energy cosmic rays events, reported by the AGASA, HiRes,
Yakutsk and Sugar collaborations, with a catalogue of galaxies of
the local universe (redshift $z\alt 0.06$). We find that the data
reproduce particularly well the clustering properties of the nearby
universe within $z \alt 0.02$. There is no statistically significant
cross-correlation between data and structures, although intriguingly
the nominal cross-correlation chance probability
 drops from ${\cal O}$(50\%) to ${\cal O}$(10\%) using the
catalogue with a smaller horizon. Also, we discuss the impact on the
robustness of the results of deflections in some galactic magnetic
field models used in the literature. These results suggest a
relevant role of magnetic fields (possibly extragalactic ones, too)
and/or possibly some heavy nuclei fraction in the UHECRs. The
importance of a confirmation of these hints (and of some of their implications) by Auger data is
emphasized.
\end{abstract}
\maketitle

\preprint{DSF/20/2007, FERMILAB-PUB-07-212-A, IFIC-07-30 }
\section{Introduction}
One of the keys towards the solution of the mysterious origin of
ultra-high energy cosmic rays (UHECRs) is the study of their
anisotropy pattern. The chances to perform (some kind of) UHECR
astronomy increase significantly at extremely high energy, in
particular due to the decreasing of deflections in the galactic
and possibly extragalactic magnetic fields. Moreover, at $E\agt
(4-5)\times 10^{19}\,$ eV the opacity of the interstellar space to
protons drastically grows due to the kinematically allowed
photo-pion production on Cosmic Microwave Background (CMB)
photons, known as the Greisen-Zatsepin-Kuzmin or GZK effect
\cite{Greisen:1966jv,Zatsepin:1966jv}. A similar phenomenon at
slightly different energies occurs for heavier primaries via
photo-disintegration energy losses. Recently, an observational
evidence for a flux suppression consistent with the GZK feature
has been reported by the HiRes collaboration \cite{Abbasi:2007sv}.
Within reasonable astrophysical assumptions, these energy-losses
phenomena impose a conservative upper limit to the distance from
which the bulk of UHECRs is emitted, of the order of a few
hundreds Mpc at most, which may enhance the chances of identifying
structures. In Ref. \cite{Cuoco:2005yd} a forecast analysis for
the Pierre Auger Observatory \cite{Auger,Abraham:2004dt} was
performed to derive the minimum statistics needed to test the
``zeroth order" hypothesis that UHECRs trace the baryonic
distribution in the universe. Assuming proton primaries, it was
found that a few hundred events at $E\agt5\times 10^{19}\,$eV are
necessary at Auger to have reasonably high chances to identify the
signature.  On the other hand, available catalogues from the
experiments of the previous generation contain ${\cal O}$(100)
events above $E\agt (4-5)\times 10^{19}\,$ eV, thus motivating a
search for possible angular patterns already in the present data
\cite{Anchordoqui:2003bx,Kachelriess:2005uf}. In particular, after
renormalizing the energy scales of the different experiments to
the HiRes one at $4\times 10^{19}$ eV, the authors of
\cite{Kachelriess:2005uf} found some evidence of a broad maximum
of the two-point autocorrelation function of UHECR arrival
directions around 25 degrees.  Since the search was made a
posteriori, the assessment of its significance is a delicate issue
and involves the determination of a penalty factor critically
dependent on the performed number of trials. The previous claim of
small scale clustering in the AGASA data and the associated debate
on its significance (see e.g.
\cite{Takeda:1999sg,Tinyakov:2001ic,Finley:2003ur}) would suggest
to take a cautionary attitude towards a posteriori claims.
However, a similar feature has been found in the Auger data {\it
alone} as well, as recently reported in \cite{Mollerach:2007vb}.
We shall thus proceed in the following under the assumption that
the signal is real, exploring some astrophysical implications.

In \cite{Cuoco:2006dx}, the present authors already tested the
qualitative interpretation of the result  (as reflecting the
large-scale structure (LSS) of UHECR sources) given in
\cite{Kachelriess:2005uf} on the light of our previous map
templates obtained from the IRAS PSCz galaxy
catalogue~\cite{saunders00a}. The observed data and the Monte
Carlo events from the catalogue share several features, which are
even more prominent if a quadratic correlation with LSS is
assumed. On the other hand, no relevant cross-correlation has been
found, which would be the smoking gun to test such scenarios.
However, this is not particularly surprising: apart for the sake
of simplicity, there is no a-priori reason to expect that cosmic
rays are 100\% made of protons, that the effects of magnetic
fields are negligible above $4\times 10^{19}\,$eV for the angular
scales considered, and that the sources trace in an unbiased way
the LSS. If some or all these assumptions are relaxed, the possibility
of a consistent scenario emerges: At ``low'' energy, both clustering and cross-correlations in UHECRs
are absent, since magnetic deflections and a very large energy-loss
horizon destroy them.
With sufficient statistics and at sufficiently ``high'' energy, cross-correlations
should eventually emerge, both because magnetic deflections scale like 1/Energy and
because of the expected shrinking of the horizon. Before this stage is reached
experimentally, it is likely
that the first hint will appear in the clustering, but not in the cross-correlation.
The reason being that the former is much more robust versus magnetic deflections
than the second one, as we shall argue.

In this paper, we extend our previous analysis in two
ways: (i) we assume a smaller horizon, i.e. biasing the
correlation with LSS towards closer sources; (ii) we study the
impact of the galactic magnetic field (GMF) on the autocorrelation
signature and on the cross-correlation signal. We anticipate that
the data reproduce particularly well the clustering properties of
the nearby universe within $z \alt 0.02$ and they are also quite
robust with respect to deflections in galactic magnetic fields. We
summarize our assumptions and techniques in Sec. \ref{sect2},
while devoting Sec. \ref{results} to present our results and attempting some
interpretations of them. In Sec.
\ref{conclusions} we briefly discuss our findings and conclude.

\section{Assumptions and methods}\label{sect2}

\begin{figure}[!tbp]
\epsfig{file=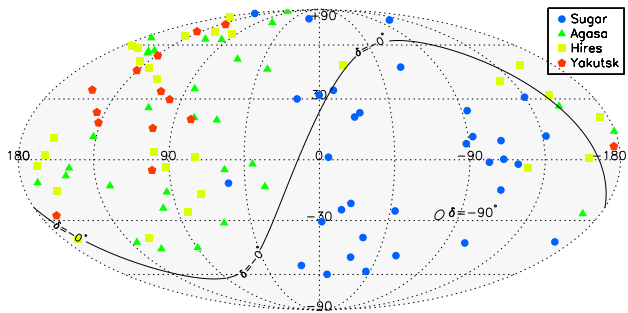,width=0.95\columnwidth} \caption{Skymap of
the UHECR arrival directions of events in galactic coordinates
with rescaled energy $E>4\times 10^{19}$~eV. The solid line is the
celestial equator.}\label{fig1}
\end{figure}

\begin{figure}[!tbp]
\vspace{1pc} \epsfig{file=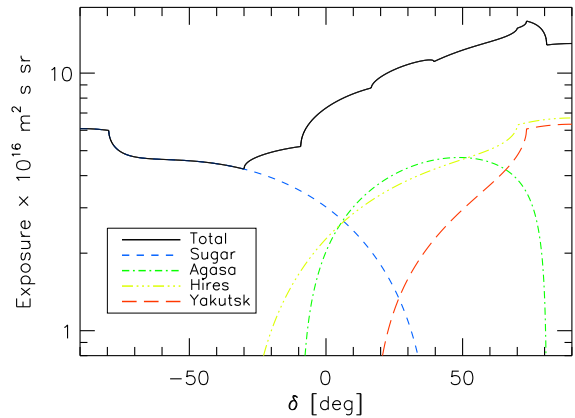,width=1\columnwidth,angle=0}
 \caption{Single and combined exposures for the
various experiment considered: Sugar (red), Hires (green), Agasa
(blue), Yakutsk (yellow), combined (black).}\label{fig2}
\end{figure}

\begin{figure}[tbp]
\begin{center}
\epsfig{file=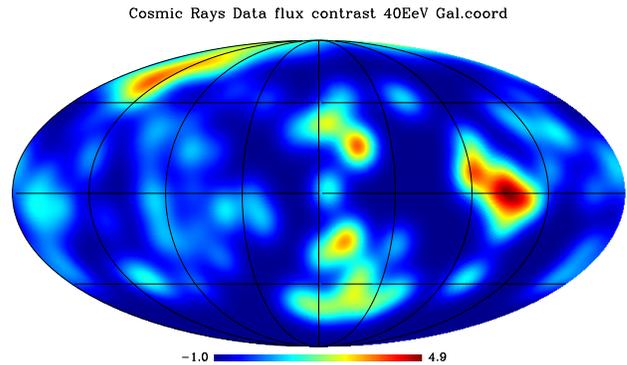,width=0.6\columnwidth,angle=90}
\end{center}
\vspace{-2pc} \caption{The UHECR flux contrast map (or excess map)
properly smoothed with a Gaussian filter of $10^\circ$
width.}\label{fig3}
\end{figure}

\subsection{The data}\label{thedata}
In our analysis, we closely follow the approach reported in
\cite{Kachelriess:2005uf,Cuoco:2006dx}, using a similar dataset
extracted from available publications or talks of the AGASA
\cite{Hayashida:2000zr}, Yakutsk \cite{yakutsk}, SUGAR
\cite{Winn:1986un}, and HiRes collaborations
\cite{Abbasi:2004ib,Hires2}. Note that we rescale {\it a priori} the
energies of the experiments to the HiRes one and consider events
above $E \geq 4\times10^{19}\,$eV in this renormalized sample. This
approach is applied hereafter in the analysis and we address the
reader to \cite{Kachelriess:2005uf} for further details. In Fig.
\ref{fig1} we show the points used in this analysis in galactic
coordinates, while Fig. \ref{fig2} reports the single and combined
exposure for the various experiments as a function of the
declination, in the limit of saturated acceptance and mediated over
the right ascension (see e.g. \cite{Sommers:2000us}). In Fig.
\ref{fig3} we show the derived UHECR excess map (flux over average
expected flux, minus one) properly smoothed by a gaussian filter of
$10^\circ$. Such a choice for the width amplitude (which has only
illustrative purposes) represents an acceptable compromise between
the few degrees of the experimental uncertainty on the arrival
direction of UHECR, and the typical angular length of the nearby
astrophysical structures of several tens of degrees. Of course, the
data have been properly weighted by the exposure.

The smoothed map in Fig. \ref{fig3} clearly shows that the most
apparent visual feature in the data is the medium scale clustering,
with the data clustered in few spots of $20^\circ$-$30^\circ$
degrees each, that in their turn are distributed almost uniformly in
the sky. This is of course the reason of the signal found in
\cite{Kachelriess:2005uf} with a proper statistical analysis. The
clustering seen in the southern hemisphere is due to the Sugar data
only and has thus a weak statistical evidence. However, the
clustering signal is indeed present and statistically significant
also considering the data from the northern hemisphere only
\cite{Kachelriess:2005uf}. Indeed, hints of this clustering in the
northern hemisphere were recognized already some years ago
\cite{Stanev:1995my}. Finally, lowering the energy threshold  the
signature disappears, excluding the possibility that the signal is
only a systematic feature coming from an incorrect modeling of the
exposures \cite{Kachelriess:2005uf}.

Concerning possible interpretations of the signal, here we report
a few general considerations, while more quantitative analyses are reported
in the following. The absence of
a correlation with the Galactic Plane or the absence of excess
toward the Galactic Center disfavor respectively Galactic astrophysical
sources and heavy relic decays in our Galactic Halo  as  origin
of these events. Qualitatively, extragalactic sources
of astrophysical nature appear the most likely accelerators. In these
scenarios, unless only a handful of sources dominate the emission, the pattern
of the arrival directions of the events should reflect to some extent
the one of large scale structures in the nearby universe.
Actually the degree of clustering observed is quite pronounced. It exceeds
also the anisotropy expected in the minimal case of proton
primaries (with a GZK horizon $z\simeq 0.06$), which is in marginal
agreement with the data (see \cite{Cuoco:2006dx} and section
\ref{results}). Thus, the few prominent structures visible
naturally suggest either a scenario where the UHE sky is dominated
by few nearby powerful sources or one where UHECRs are
produced by a relatively larger number of sources significantly
biased with overdensities in the local universe (within $z\simeq
0.02$). Both scenarios require an important role of magnetic
fields, either galactic or extragalactic, to accommodate deflections
of the order $\agt$10$^\circ$.  In the former case such deflections
are necessary to explain the large smearing of the
point sources emission and in the latter to justify the lack of a significant
correspondence between the data and the nearby galaxy clusters.

\subsection{The models}\label{themodels}
The previous discussion motivates an extension of the previous
analysis reported in \cite{Cuoco:2006dx} along two directions:
\begin{itemize}
\item[($I$)]  assuming a smaller
horizon, i.e. biasing the correlation with LSS towards closer
sources;
\item[($II$)] studying the impact of the GMF on the
autocorrelation signature and on the cross-correlation signal.
\end{itemize}
Both extensions should be regarded as ``first order" refinements
of our previous study. The point ($II$) does not need much
justification: it is important to establish the robustness of the
previous results with respect to the effects of astrophysical
magnetic fields. Even if extragalactic magnetic fields may have a
major role in shaping or preserving the UHECR anisotropies, very
little is known about them (see \cite{Dolag:2003ra} and
\cite{ems}). On the other hand we know for sure that a regular GMF
exists---although we have only rough ideas on its magnitude and
structure---and it is in principle relevant for UHECR deflections,
even in the case of pure proton composition. In the following, we
shall then consider how our results change when data are corrected
for the effects of a few GMF models available in the literature.
In particular, we shall use the three models HMR, TT, and PS
employed in \cite{Kachelriess:2005qm}, which we refer to for
details. To account for the deflections in the GMF in our analyses
we shall follow the  back-tracking technique described in
\cite{Takami:2005ij,Harari:1999it}. The technique consists in
mapping the arrival CRs directions on the Earth backward outside
the GMF to obtain a map of the GMF deflections. We then apply the
mapping to the extragalactic expected CR map and correct it for
the GMF displacements.  The extragalactic CRs map
$F({E_{\rm{cut}}},\hat{\Omega})$ expected at an energy greater
than $E_{\rm{cut}}$ at the direction $\hat{\Omega}$ is obtained as
described in \cite{Cuoco:2005yd}. The map is then convolved with
the GMF deflections to have
$F({E_{\rm{cut}}},\hat{\Omega}(\hat{\Omega}'))$ where
$\hat{\Omega}(\hat{\Omega}')$ is the mapping produced by the
back-tracking technique. This method is fully suited for the cases
in which energy losses along the particle track are negligible and
when the particle energy  is large enough to exclude loops and/or
trapped regions during the propagation. Both these conditions are
satisfied for the UHECRs and for the GMFs we considered. Also note
that an isotropic sky remains isotropic under the GMF
transformation, in agreement with the expectation from the
Liouville theorem. We refer to
\cite{Kachelriess:2005qm,Takami:2005ij,Harari:1999it} for details.

For simplicity we only consider the mapping produced for a fixed
rigidity corresponding to the energy $E_{\rm{cut}}$ (with the
choice $E_{\rm{cut}}$=40 EeV  in the present case). This should be
a reasonable approximation, equivalent to replace the steep
($\propto E^{-3}$) UHECRs spectrum above $E_{\rm{cut}}$  with a
delta-function at $E_{\rm{cut}}$. Beside the steepness of the
UHECR spectrum, a further motivation for this approximation is
that we are not considering the shift of single objects but of an
overall map, already smoothed at a scale of order $\sim 5^\circ$
(we are not interested to the very small scales, indeed). Finally,
we shall show in the next section that the auto-correlation
analysis is quite insensitive to the details of the GMFs or the
assumed rigidity as long as the magnetic deflections effects
remains moderate, so that the approximation is also  justified a
posteriori, at least for autocorrelations studies. The effects are
potentially larger for cross-correlation analyses, which however
are already less robust for  other reasons.

A further possible problem is given by the fact that the mask
region present in the catalogue and excluded from the analysis is
distorted by the effect of the GMF, so that in principle one
should exclude, case by case, the regions which the mask is mapped
into by the GMF. We neglect this effect assuming the the mask is
approximately mapped in itself by the GMF transformation. This is
a quite good approximation for the region near the galactic plane
while it is not satisfied by the two narrow stripes. However, the
the stripes amount to about 10\% of the total mask and only
roughly 2\% of the whole sky, which is a very small bias for our
purposes in this work.

\begin{figure*}[tp]
\begin{center}
\begin{tabular}{cc}
\epsfig{file=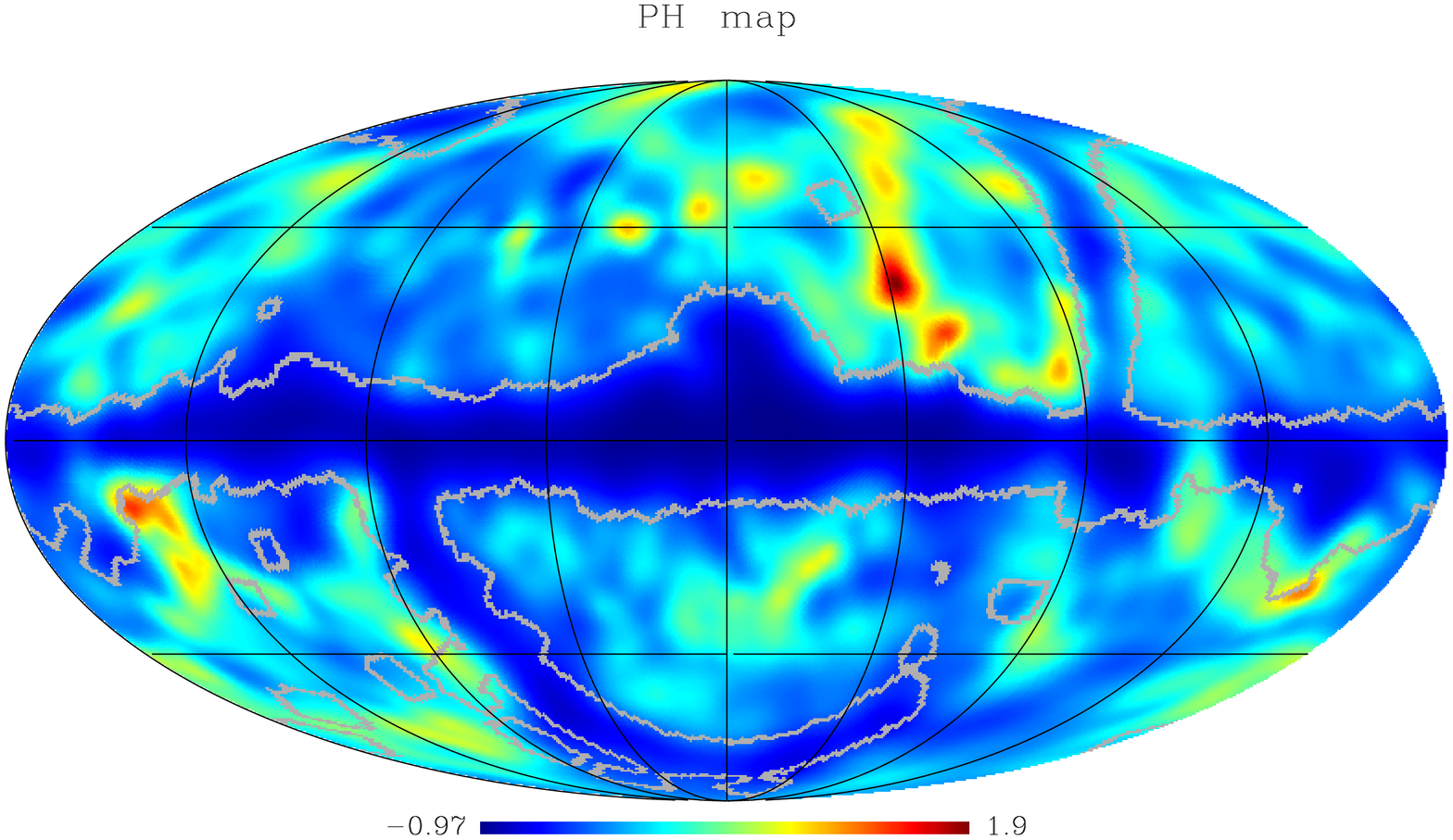,width=4.7cm,angle=90} & \hspace{1pc}
\epsfig{file=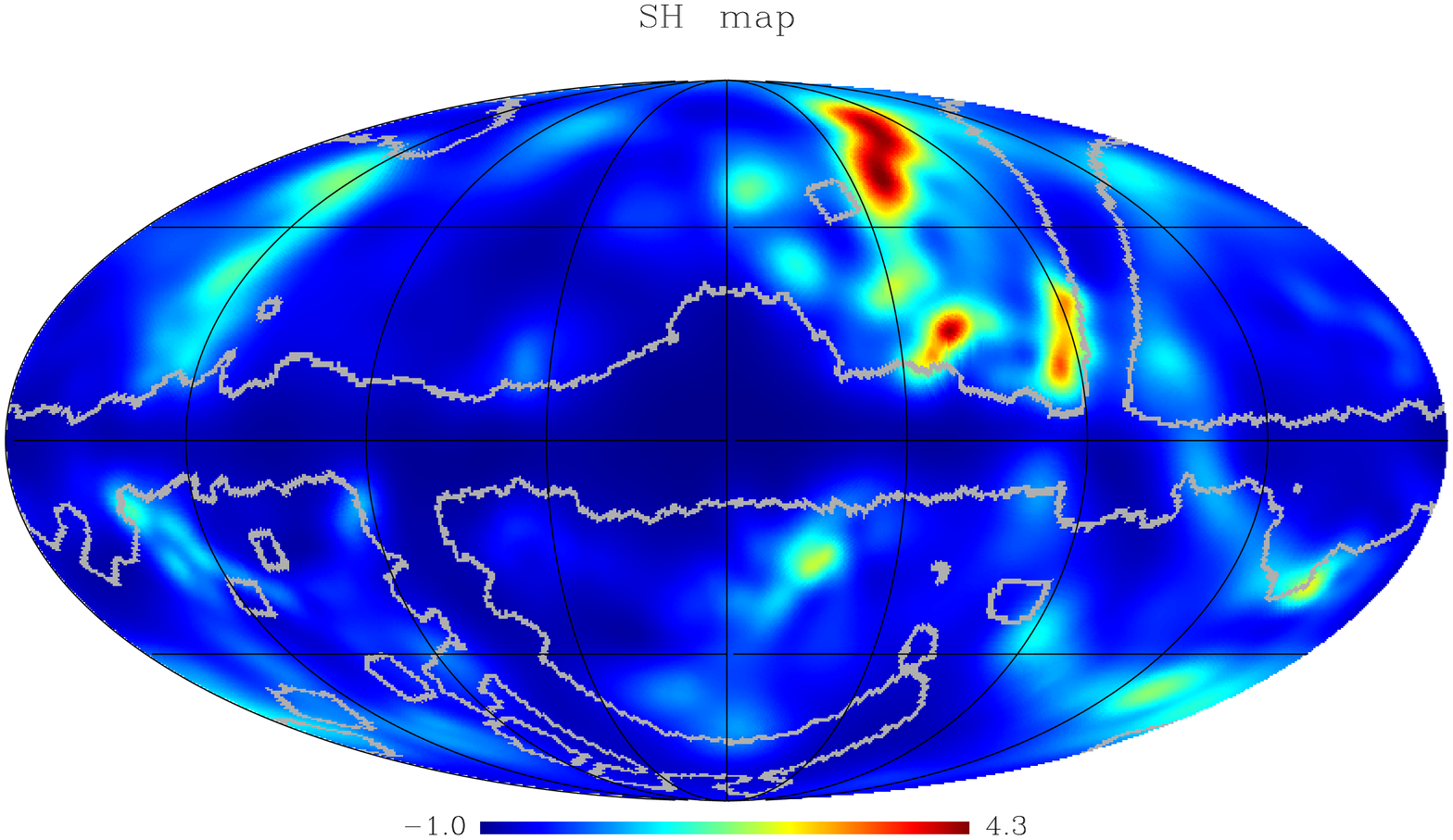,width=4.7cm,angle=90}\\
\epsfig{file=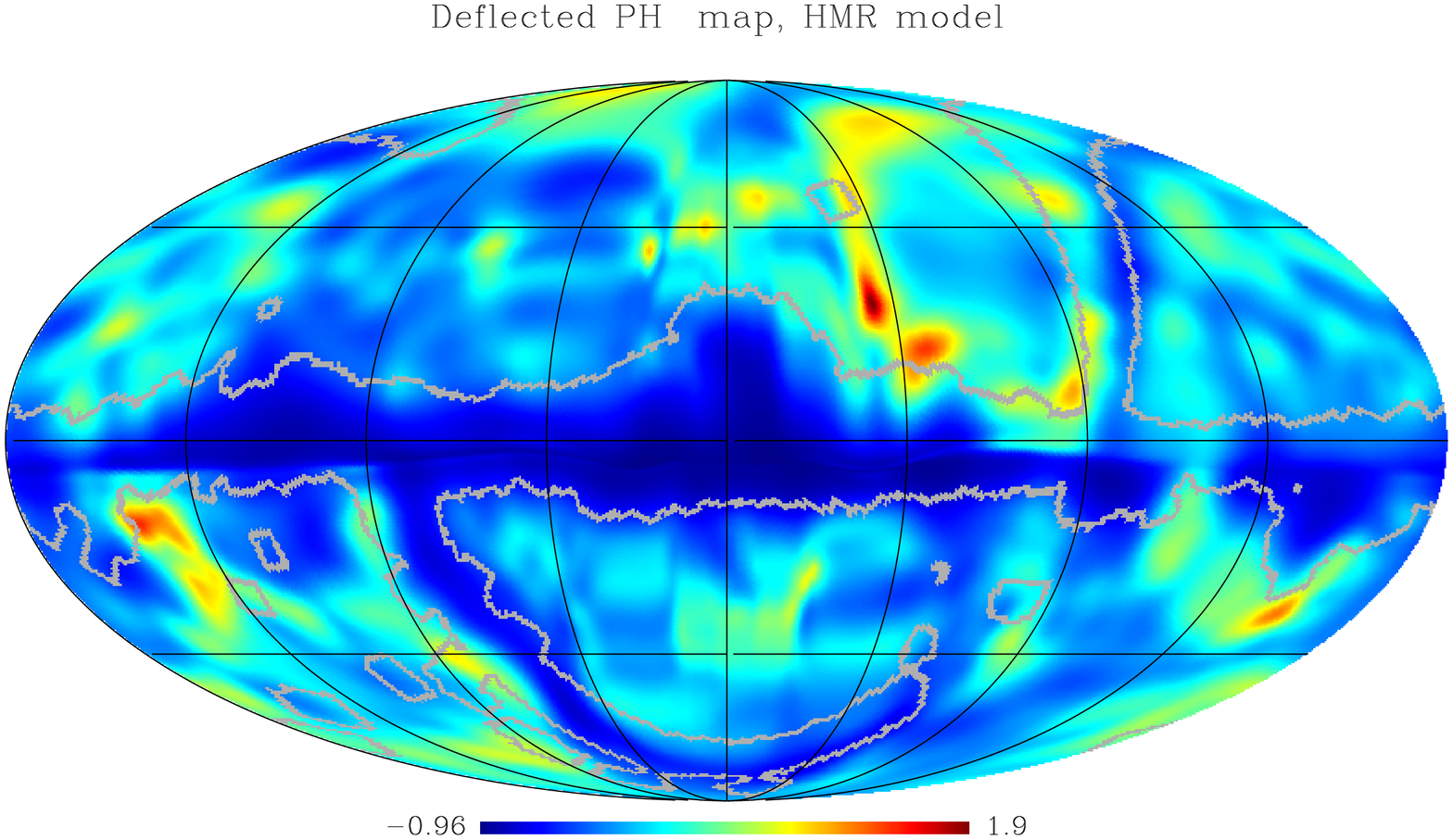,width=4.7cm,angle=90} & \hspace{1pc}
\epsfig{file=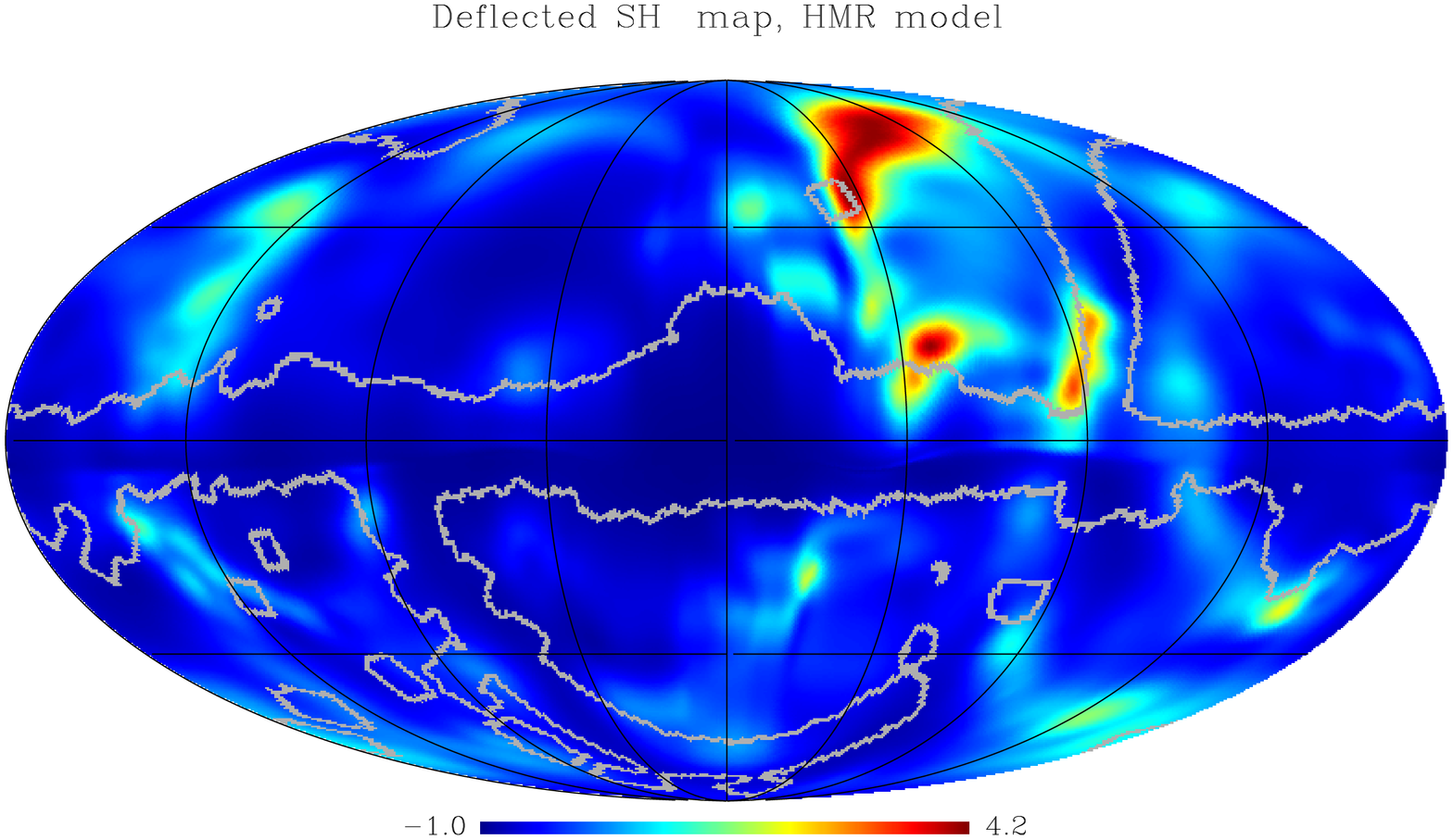,width=4.7cm,angle=90} \\
\end{tabular}
\end{center}
\caption{Top row: Excess maps of PH (left) and SH models (right)
in galactic coordinates. The grey contour bounds the blind region
of PSCz catalogue. Bottom row: Galactic excess maps of the PH
(left) and SH models (rigth) taking into account the Galactic
magnetic field correction for a specific model (HMR in
\cite{Kachelriess:2005qm}) and assuming proton primaries ($Z=1$).
\label{fig4}}
\end{figure*}

From a qualitative point of view, the point ($I$) is reasonable,
too. In many scenarios, only relatively nearby sources (if any)
may be identifiable in cosmic ray maps. There are several
plausible reasons for that. Even in absence of magnetic fields, an
heavier composition implies a different energy-loss horizon for
UHECRs \cite{Harari:2006uy,Hooper:2006tn}. For the energy
threshold considered here ($E\agt 4\times 10^{19}\,$eV), this is
smaller than the proton one. In presence of extragalactic magnetic
fields, the propagation of a UHECR may greatly differ from a
straight line and in principle may even happen in a diffusive
regime
\cite{Sigl:1998dd,Lemoine:1999ys,Deligny:2003rp,Parizot:2004wh}.
Although it is unlikely that the propagation is truly diffusive,
Gpc-scale pathlengths for protons injected within a few hundreds
Mpc may be common even above 10 EeV \cite{Armengaud:2004yt}. A
non-negligible role of magnetic fields would have two
consequences: for a given energy-loss mechanism, it is clear that
the true horizon may be significantly shorter than the expected
one. Thus, UHECRs above a given $E_{\rm th}$ may be largely
collected within a region smaller than the linear energy-loss
horizon. More important, apart for energy losses, the longer the
propagation time, the smaller the chance that intrinsic
anisotropies may survive (in some form). Finally, since UHECR
source likely have to meet special accelerator requirements, it is
reasonable to conceive a relatively rare population of sources,
possibly strongly biased with respect to LSS.

However, how to implement in practice point $(I)$ is admittedly {\it
not} model independent. One possibility may be to cut arbitrarily a
LSS catalogue to some redshift $z_{\rm cut}$, and consider only
correlations with structures within this distance, assuming for the
rest that UHECRs are unbiased tracers of LSS (i.e. neglecting
otherwise energy loss effects). Another possibility is to create
anisotropy map templates of specific scenarios for UHECR
composition, sources, and extragalactic magnetic fields, comparing
them with the observed configurations of data in order to infer the
best model. Although this will be the way to proceed when high
statistics will be achieved, at the moment it could just dilute the
basic consequence of our assumption ($I$) under a large number of
unknown parameters. To keep some physical-inspired input in a {\it
toy model}, we shall compare the distribution of data as in Fig.
\ref{fig1} with the LSS maps obtained by convolution of the PSCz
catalogue with an energy-loss window function corresponding to
protons twice more energetic, i.e. $E = 8\times 10^{19}\,$eV,
implying an effective horizon $ z\simeq 0.02$ \cite{Cuoco:2005yd}.
We shall denote this scenario as ``small horizon" (SH), as opposed
to the ``proton horizon" (PH) as treated in \cite{Cuoco:2006dx} and
corresponding to the minimal assumption of protons primaries with $E
\geq 4\times 10^{19}\,$eV propagating in a negligible EGMF (usual
GZK horizon $z\simeq 0.06$). In the top two panels of Fig.
\ref{fig4} we report the PH and SH maps. The smoothing is variable
and it is related to the adaptive smoothing applied to the PSCz
catalogue to minimize the effect of the shot noise. We emphasize
that this should be considered a toy model, and not a realistic
scenario for UHECR sources or composition. However, our toy model
may be indicative of a plausible situation where at least for
anisotropy searches the useful UHECR horizon is relatively short.

\subsection{Statistical tools}\label{thetools}
For the statistical analysis we define the (cumulative)
autocorrelation function $w$ as a function of the separation angle
$\delta$ as
\begin{equation}
w(\delta)=\sum_{i=2}^{N_d}\sum_{j=1}^{i-1}\Theta(\delta-\delta_{ij}),
\end{equation}
where $\Theta$ is the step function, $N_d$ the number of CRs
considered and $\delta_{ij} =\arccos(\cos\rho_i\cos\rho_j +
\sin\rho_i \sin\rho_j \cos(\phi_i -\phi_j))$ is the angular distance
between the two cosmic rays $i$ and $j$ with coordinates $(\rho,
\phi)$ on the sphere. Analogously, one can define the correlation
function $\xi(\delta)$ as
\be \xi(\delta) = \sum_{i=1}^{N_d}\sum_{a=1}^{N_s}
\Theta(\delta-\delta_{ia}) \,, \ee
where $\delta_{ia}$ is the angular distance between the CR $i$ and
the candidate source $a$ and $N_s$ is the number of source objects
considered.

\begin{figure*}[!t]
\epsfig{file=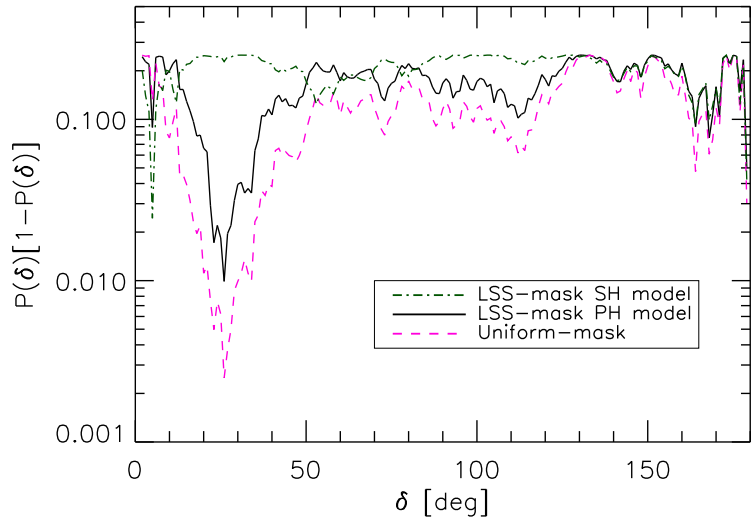,width=1.25\columnwidth} \caption{Chance
probability of auto correlation taking as reference model an uniform
distribution, the linear correlation model of \cite{Cuoco:2006dx}
(LSS-PH model) and the presently considered model with a smaller
horizon (LSS-SH model). \label{fig5}}
\end{figure*}

We perform a large number $M\simeq 10^5$ of Monte Carlo simulations
of $N$ data sampled from a distribution on the sky corresponding to
the hypothesis $H$ (e.g., uniform, LSS, etc.) and for each
realization $j$ we calculate the autocorrelation function
$w_j^{H}(\delta)$. The sets of random data match the number of data
for the different experiments passing the cuts after rescaling, and
are spatially distributed according to the exposures of the
experiments.
 The formal
probability $P^H(\delta)$ to observe an equal or larger value of
the autocorrelation function by chance is
\begin{equation}
P^{H}(\delta)=\frac{1}{M}\sum_{j=1}^M\Theta[w_j^{H}(\delta)-w_\star(\delta)],\label{Pdelta}
\end{equation}
where $w_\star(\delta)$ is the observed value for the cosmic ray
dataset and the convention $\Theta(0)=1$ is being used. Relatively
high values of $P$ {\it and} $1-P$ indicate that the data are
consistent with the null hypothesis being used to generate the
comparison samples, while low values of $P$ {\it or} $1-P$ indicate
that the model is inappropriate to explain the data. That is, in the
following we shall plot the function $P(\delta)\times
[1-P(\delta)]$, which vanishes if any of $P$ or $1-P$ vanishes and
has the theoretical maximum value of $1/4$. Thus, the higher its
value is the more consistent the data are with the underlying
hypothesis. Note also that by construction the values at different
$\delta$ of the function $P(\delta)$ are not independent.

To calculate the cross-correlation probability, we perform a large
number $M(\simeq 10^5)$ Monte Carlo realization of $N$ events is
sampled according to the LSS probability distribution, and for
each realization $i$ we calculate the function $\xi_i^{\rm
LSS}(\delta)$. We generate analogously $M$ random datasets from an
uniform distribution, and calculate $\xi_j^{\rm uni}(\delta)$. We
have thus $M^2$ independent couples of functions $(i,j)$. The
fraction of the $M^2$ simulations where the condition $\xi^{\rm
uni}(\delta)\geq \xi^{\rm LSS}(\delta)$ is fulfilled is the
probability
\begin{equation}
P_{\xi}(\delta)=\frac{1}{M^2}\sum_{i=1}^M\sum_{j=1}^M\Theta[\xi_i^{\rm
uni}(\delta)-\xi_j^{\rm LSS}(\delta)]\,.
\end{equation}

A technical detail  of the analysis  is related to the presence of
the catalogue mask. This includes a zone centered on the galactic
plane and caused by the galactic extinction and a few, narrow
stripes which were not observed with enough sensitivity by the
IRAS satellite. These regions are excluded from our analysis with
the use of the binary mask available with the PSCz catalogue
itself. This reduces the available sample by about 10\%.

\begin{figure*}[!htbp]
\begin{center}
\begin{tabular}{cc}
\epsfig{file=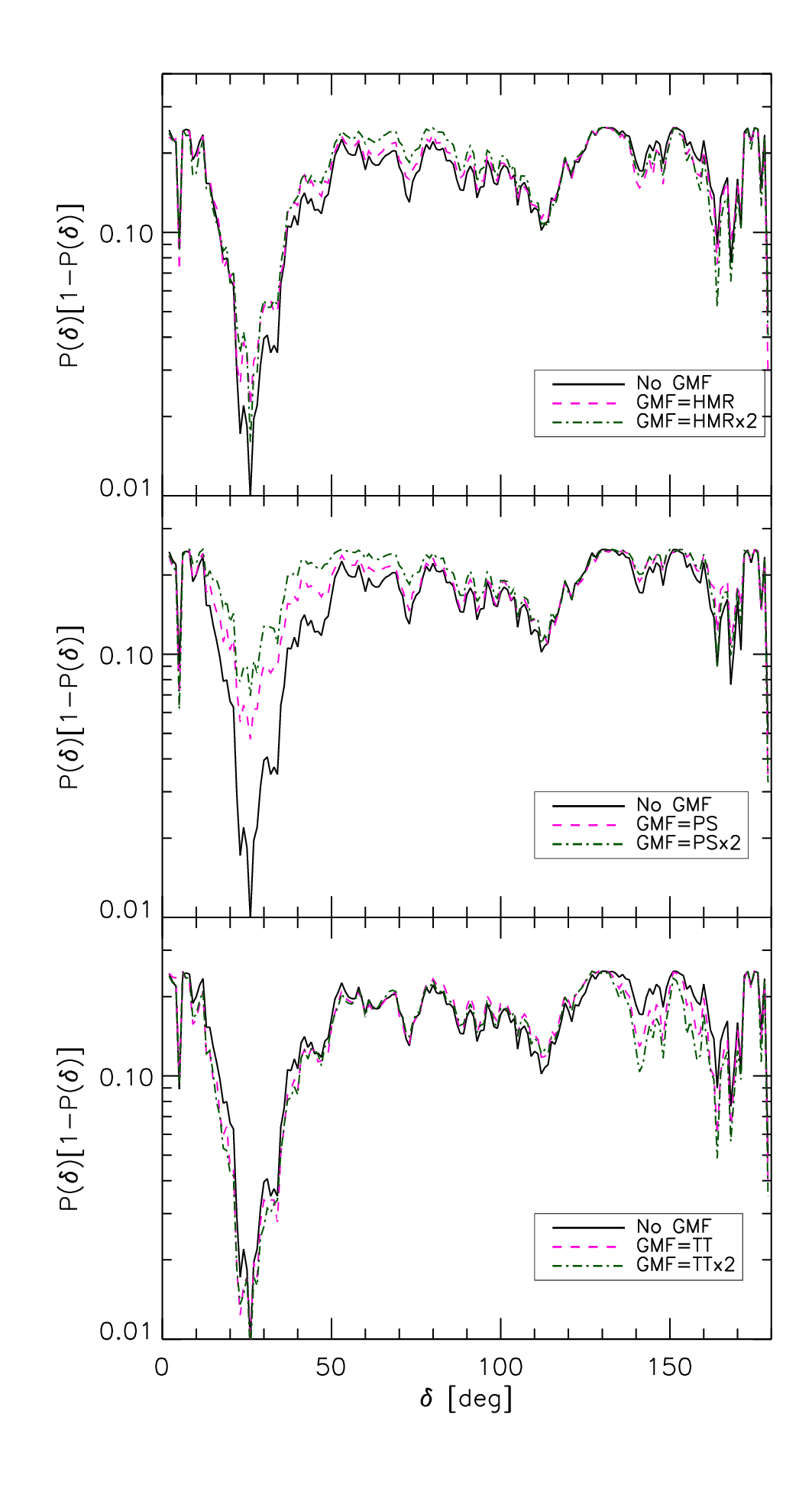,width=1.0\columnwidth,angle=0} &
\hspace{1pc}
\epsfig{file=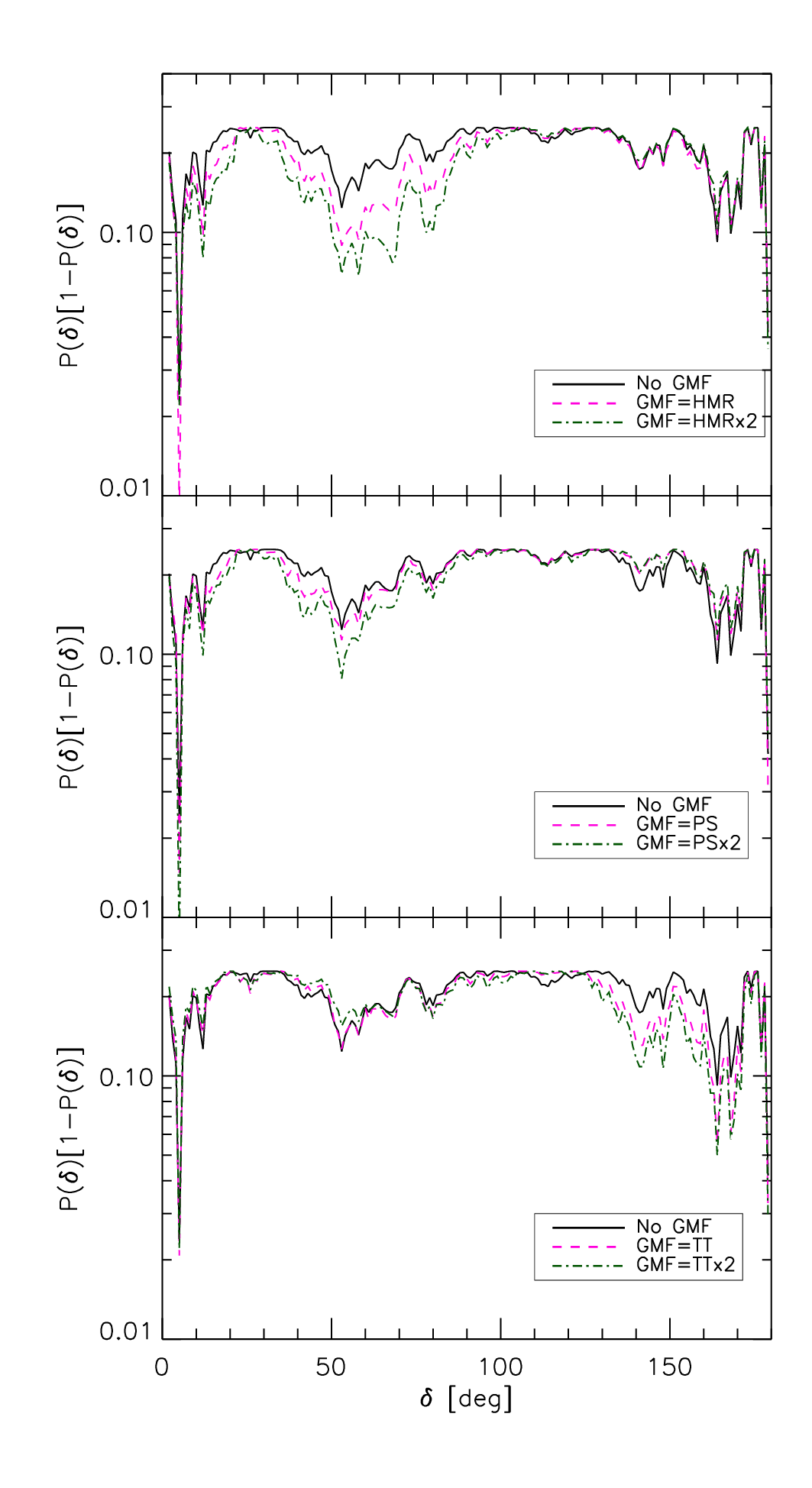,width=1.0\columnwidth,angle=0}
\end{tabular}
\end{center}
\vspace{-2.5pc} \caption{Chance probability of autocorrelation of
the data for different GMF models and for the PH model maps (left)
and SH model maps (right) taken as reference sample. The cases
$Z=1$ and $Z=2$ are shown in each panel.\label{fig6}}
\end{figure*}

\section{Results}\label{results}

By repeating our analysis in \cite{Cuoco:2006dx} following the SH
model and without considering for the moment the effects of the GMF,
we obtain the results shown in Fig. \ref{fig5}. The SH model seems
to explain extremely well the clustering properties of the data,
with the related $P\times(1-P)$ curve almost coincident with the
ideal $P=0.5$ expectation. Not surprisingly, this can be understood
after a visual inspection of the maps in Fig. \ref{fig3} (data) and
Fig. \ref{fig4} (models). While the map from protons with $E \geq
4\times 10^{19}\,$eV (the PH model) is still too much isotropic with
respect to the data, in the SH map the number of clusters and their
distribution resemble much more the data, in that it leaves typical
``voids" between clustered hot-spots observed.

\begin{figure}[!htp]
\epsfig{file=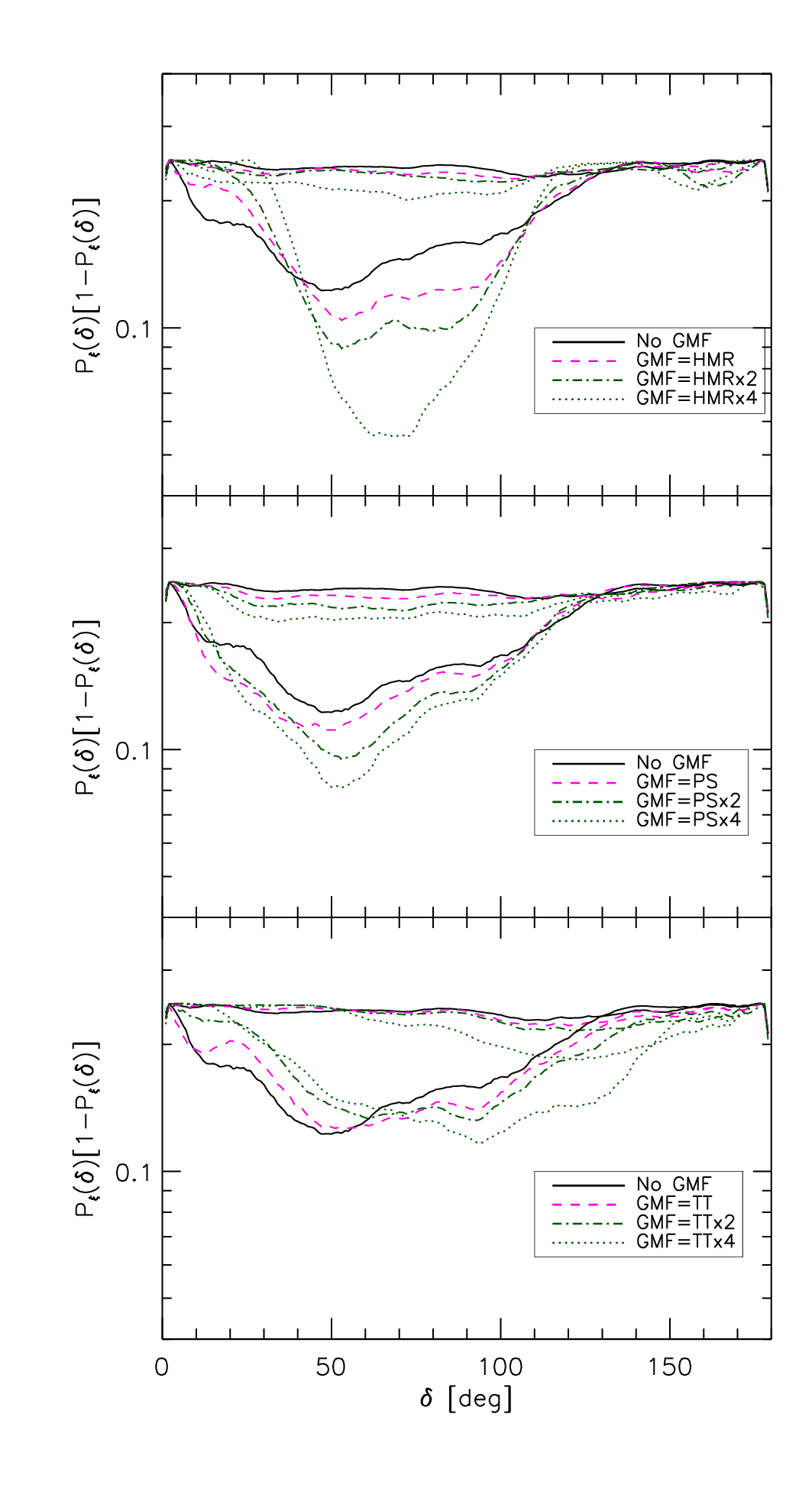,width=1.0\columnwidth,angle=0} \vspace{-2.0pc}
\caption{Cross-correlation chance probability between the different
LSS models and the data, with the solid lines representing the
result uncorrected for the GMF. Upper set of curves in each panel
refer to the PH model while lower ones to the SH model.
\label{fig7}}
\end{figure}

Our next step is to investigate the effects of the galactic
magnetic field. In the two bottom panels of Fig. \ref{fig4} we
show the effective modification of LSS structures happening for
the PH and the SH scenario, assuming as example the HMR model in
\cite{Kachelriess:2005qm}. In general, besides the shifting of the
positions of the structures, as expected the GMF introduces in the
deflected maps also other peculiar lensing phenomena like shearing
and (de)magnification \cite{Harari:1999it}. More quantitatively,
the effects of the GMF are studied in the following through the
modifications induced in the auto and cross-correlation functions.
In Fig. \ref{fig6} we investigate the effect of the GMF on the
signature in the autocorrelation function. We also show the effect
of changing the rigidity of the particles. Actually, the equations
of motion for CRs in the GMF only depend on the parameter
$C=B\times Z/E$ where $B,Z,E$ are respectively the GMF
normalization, the particle atomic number (electric charge of the
nucleus) and the particle energy.  A combination of parameters
that leaves unchanged $C$ is thus completely degenerate from the
point of view of propagation in the GMF (though, of course, not
for the energy losses in the propagation in the extragalactic
sky.) As a general consideration, we see that at least for the
baseline cases considered, the correction for the GMF does not
destroy the pattern in $P(1-P)$, but can improve or worsen it at
most by a factor of a few. On the other hand, extreme changes in
$C$ may significantly alter the pattern of the function.

Finally, in Fig. \ref{fig7} we report the results of the cross
correlation analysis.  Note that, while in the previous case we
were repeating the same test of Ref. \cite{Kachelriess:2005uf},
here we perform a different test, and to assess the confidence
level of any statistically significant signal we might find one
should carefully evaluate the penalty factor. Unfortunately, in no
case we find a statistically significant signal (namely, not even
at the nominal level). Yet, qualitatively all the SH maps show
some improvement with a nominal $P_{\xi}\sim 10\%$ (with respect
to $P_{\xi}\sim 50\%$ for the PH case), even without use of the
GMF correction. This is understood since we have about the same
number of clusters in the data and in the map and typically one
can find a correspondence between the two within a radius of
roughly $50^\circ$. A more significant signal of cross-correlation
should eventually peak within $\sim 25^\circ$ (the typical size of
the clusters) signaling a superposition of the data and map
clusters. Once again, no significant difference arises when the
data are corrected for the GMF. Although the minimum of the
probability can change by up to a factor of a few, it does not
move towards $\delta\simeq 0^\circ$, as it should be if the GMF
were correctly shifting the hot-spots.

An important point to stress is that
while an evidence of cross-correlation
would be tantalizing signature of  a discovery, the lack of it
can not be easily used as an argument against the hypothesis.
The cross-correlation signal is indeed much more sensitive than the
autocorrelation one to magnetic fields deflection and, importantly, to unknown
experimental systematic effects.  For the case at hand, the
main responsables of the displacement up to 50 degrees between
clusters in the data and overdensities in the LSS are the hot-spots from the SUGAR
data, which is the experiment among the ones considered which mostly
suffers for a poor angular resolution, beside not
well-understood systematics in the energy scale determination.
Indeed, limiting the analysis to the northern hemisphere
experiments only, the data show quite a good cross-correlation
with the local over-density of matter, especially
within $z\sim$ 0.02. More noticeably, they fall relatively close to
the so-called Super-Galactic plane, as can be appreciated also from a
comparison between Fig.~\ref{fig1} and Fig.~\ref{fig4}. This
correspondence was already noticed by the authors of ref.~\cite{Stanev:1995my} and further assessed in
ref.~\cite{Burgett:2003yg}. So, while the features in the cross-correlation
function vary quite a bit excluding e.g. one dataset, the auto-correlation
ones do not, as discussed more extensively in~\cite{Kachelriess:2005uf}.

Aware of this caveat, it  is still worth exploring the consequences of
assuming that displacements up to 50$^\circ$ with respect to the true
sources are effective. Under this
hypothesis, and, if the signal corresponds to extragalactic
structures, we would be brought to conclude that: (i) if UHECRs are
dominated by protons, then there are significant deflections
by extragalactic magnetic fields. Indeed, although the GMF may be
not well reproduced by current models, even changing within
reasonable ranges the GMF geometry and intensity no appreciable
cross-correlation at small angles appears. (ii) If there is a
significant fraction of heavy nuclei in the UHECR flux, results
may be also explained with a negligible role of extragalactic
magnetic fields, attributing to GMF deflections the significant
($\sim 30^\circ$--$50^\circ$) displacement between the observed
clusters in the data and the real galaxy clusters.  Note that, if
in any case overall deflections as large as $\sim 50^\circ$ are
effective, peculiar manifestations of regular deflections in the
magnetic field---like elongations directed towards structures with
a proper ordering of energies---may not be observable due to
non-negligible non-linearities, especially in the case of a
chemically inhomogeneous sample of UHECRs. Yet, if this interpretation
would turn out to be the correct one, we expect that once higher
statistics will be available {\it at higher energies}, these
features will eventually show up in the data, a prediction that
can eventually be confirmed by Auger.

\section{Discussion and Conclusion}\label{conclusions}
The anisotropy pattern of the combined UHECRs data (re-scaled energy $E=4\times 10^{19}\,$eV in HiRes
scale), although compatible with isotropy at very large angular scales, shows a
peculiar medium scale clustering corresponding to 6-7 spots of
roughly $20^{\circ}$-$30^{\circ}$ degrees of extension distributed
uniformly in the sky. If confirmed, this would have a
wealth of consequences for the long-awaited astronomy of UHECRs.
At a general level, the absence of
a correlation with the Galactic Plane or the absence of an excess
toward the Galactic Center disfavor respectively Galactic astrophysical
sources and heavy relic decays in our Galactic Halo  as  origin
of these events. Extragalactic sources
of astrophysical nature appear instead the most likely accelerators consistent
with these features. Should forthcoming data show similar features,
the first realistic quantities one could extract from the clustering are constrains the number density
of the sources as well as their type, from their bias with respect to LSS  (see
\cite{Cuoco:2007id} and references therein). As next goal, one should be able to
establish explicit cross-correlation with extragalactic structures and/or hints in that
direction, as elongations of events directed toward the sources, whose distance should scale
inversely to the UHECR energy. More ``conventional"
astronomy, determining the locations of single UHECR sources and perhaps of
their spectrum is likely demanded to a subsequent phase when sufficient
statistics will be available.

Although the clustering properties in the data are intriguing, the interpretation of the signature is puzzling, especially in
absence of a significant statistics at higher energy and of chemical
composition constraints. The comparison between significant sets of
data at different energy cuts may reveal the importance of magnetic
deflection effects. At the moment, we can only speculate on the
possible implications of the signal---assuming it is not a
statistical fluke---under some simplifying hypotheses.

One possibility is that these excesses may trace LSS overdensities
in the near universe (within the GZK-sphere). The autocorrelation
analyses reported in this paper show that this interpretation is
indeed favored in particular if the effective horizon is smaller
than the GZK one for protons of the assumed energy. Both a
significant fraction of heavier nuclei and a significant role of
extragalatic magnetic fields may cause this effect (the former
 might be favored by recent Auger data \cite{Unger:2007mc}.)
Although not statistically significant, this interpretation may be
supported by a weak hint of a broad minimum in the cross
correlation function (at the level of nominal chance probability
of 10\%-15\%) around 50$^\circ$ if a small horizon ($z\alt 0.02$)
is assumed. Both signatures are relatively robust with respect to
deflections in typical GMF models, although some marginal
improvement or worsening may arise for some choices of the GMF
model and effective rigidities. In this case, the size
of the hot-spots would be due partly to the one of the largest overdensities
in the local LSS and partly to magnetic smearing needed to
explain the overall deflection with respect to the LSS. The latter effect would be in
general subleading but for the SUGAR hotspots in the Southern Emisphere, which
are the most distant ones from overdensities.
This may be physically  associated to the more intense magnetic fields towards the
central regions of our Galaxy to which SUGAR is pointing. An alternative interpretation of
the data is that they are due to very few ($\cal{O}$(5-6)) powerful sources. Yet,
the smearing of a point-like emission to the level of the observed
spots of $\cal{O}$(20$^\circ$) would require a quite extreme magnetized
environment \cite{Sigl:1998dd,Lemoine:1999ys}.

In any case, the hints for some structures in the data are very
exciting, and we urge an independent cross-check with the nowadays
large statistics collected by Auger. If confirmed, together with the
indication for the presence of a GZK-like feature in the energy
spectrum of HiRes data \cite{Abbasi:2007sv}, this likely implies
that UHECR are dominated by astrophysical sources (as opposed to
exotic scenarios). However, far from being the end of the UHECR
saga, the combined use of spectral information, chemical composition
constraints, and anisotropy maps at different energies would offer
the tools for the long-awaited hunt for the UHECR accelerators.

\section*{Acknowledgments}
We thank M. Kachelrie{\ss} and D. Semikoz for comments. Use of the
publicly available Healpix  package \cite{Gorski:2004by} is
acknowledged.   P.S. acknowledges support by the US Department of
Energy and by NASA grant NAG5-10842. G.M. acknowledges support by
Generalitat Valenciana (ref. AINV/2007/080 CSIC). This work was
also supported by PRIN06 ``{\it Fisica Astroparticellare}'' by
Italian MIUR.




\begin{thebibliography}{00}
\bibitem{Greisen:1966jv}
  K.~Greisen,
   ``End To The Cosmic Ray Spectrum?,''
  Phys.\ Rev.\ Lett.\  {\bf 16}, 748 (1966).

\bibitem{Zatsepin:1966jv}
  G.~T.~Zatsepin and V.~A.~Kuzmin,
  ``Upper Limit Of The Spectrum Of Cosmic Rays,''
  JETP Lett.\  {\bf 4}, 78 (1966)
  [Pisma Zh.\ Eksp.\ Teor.\ Fiz.\  {\bf 4}, 114 (1966)].

\bibitem{Abbasi:2007sv}
  R.~Abbasi {\it et al.}  [HiRes Collaboration],
  ``Observation of the GZK cutoff by the HiRes experiment,''
  astro-ph/0703099.

\bibitem{Cuoco:2005yd}
  A.~Cuoco, R.~D.~Abrusco, G.~Longo, G.~Miele and P.~D.~Serpico,
``The footprint of large scale cosmic structure on the ultra-high
energy cosmic ray distribution,''
  JCAP {\bf 0601} (2006) 009
  [astro-ph/0510765].

  \bibitem{Auger} Pierre Auger Collaboration, 1996 {\it
The Pierre Auger Project Design Report}, FERMILAB-PUB-96-024.

\bibitem{Abraham:2004dt}
Abraham J {\it et al.} (Pierre Auger Collaboration), ``Properties
and performance of the prototype instrument for the Pierre Auger
Observatory,'' 2004 {\it Nucl.\ Instrum.\ Meth.\ A} {\bf 523} 50.


\bibitem{Anchordoqui:2003bx}
  L.~A.~Anchordoqui, C.~Hojvat, T.~P.~McCauley, T.~C.~Paul, S.~Reucroft, J.~D.~Swain and A.~Widom,
  ``Full-sky search for ultrahigh-energy cosmic ray anisotropies,''
  Phys.\ Rev.\  D {\bf 68} (2003) 083004
  [astro-ph/0305158].


\bibitem{Kachelriess:2005uf}
  M.~Kachelrie{\ss} and D.~V.~Semikoz,
``Clustering of ultra-high energy cosmic ray arrival directions on
medium scales,''
  Astropart.\ Phys.\  {\bf 26} (2006) 10
  [astro-ph/0512498].

\bibitem{Takeda:1999sg}
  M.~Takeda {\it et al.},
``Small-scale anisotropy of cosmic rays above 10**19-eV observed
with  the Akeno Giant Air Shower Array,''
  Astrophys.\ J.\  {\bf 522}, 225 (1999)
  [astro-ph/9902239].

\bibitem{Tinyakov:2001ic}
  P.~G.~Tinyakov and I.~I.~Tkachev,
``Correlation function of ultra-high energy cosmic rays favors point
  sources,''
  JETP Lett.\  {\bf 74}, 1 (2001)
  [Pisma Zh.\ Eksp.\ Teor.\ Fiz.\  {\bf 74}, 3 (2001)]
  [astro-ph/0102101].

\bibitem{Finley:2003ur}
  C.~B.~Finley and S.~Westerhoff,
``On the evidence for clustering in the arrival directions of
AGASA's ultrahigh energy cosmic rays,''
  Astropart.\ Phys.\  {\bf 21}, 359 (2004)
  [astro-ph/0309159].


\bibitem{Mollerach:2007vb}
  S.~Mollerach [Pierre Auger Collaboration],
  ``Studies of clustering in the arrival directions of cosmic rays detected
  at the Pierre Auger Observatory above 10 EeV,''
  arXiv:0706.1749 [astro-ph].

\bibitem{Cuoco:2006dx}
  A.~Cuoco, G.~Miele and P.~D.~Serpico,
  ``First hints of large scale structures in the ultra-high energy sky?,''
  Phys.\ Rev.\  D {\bf 74} (2006) 123008
  [astro-ph/0610374].

\bibitem{saunders00a}
  W.~Saunders {\it et al.},
  ``The PSCz Catalogue,''
  Mon.\ Not.\ Roy.\ Astron.\ Soc.\  {\bf 317}, 55 (2000).
 [astro-ph/0001117].

\bibitem{Hayashida:2000zr}
  N.~Hayashida {\it et al.},
``Updated AGASA event list above 4$\times$10$^{19}$ eV,''
astro-ph/0008102.

\bibitem{yakutsk} Talk of M. Pravdin at the 29$^{\rm th}$ ICRC Pune 2005,\\
\texttt{http://icrc2005.tifr.res.in/htm/PAPERS/HE14/
rus-pravdin-MI-abs1-he14-poster.pdf}

\bibitem{Winn:1986un}
  M.~M.~Winn {\it et al.},
``The cosmic ray energy spectrum above 10$^{17}$ eV,''
  J.\ Phys.\ G {\bf 12} (1986) 653.

\bibitem{Abbasi:2004ib}
  R.~U.~Abbasi {\it et al.}  [The High Resolution Fly's Eye Collaboration
                  (HIRES)],
``Study of small-scale anisotropy of ultrahigh energy cosmic rays
observed in stereo by HiRes,''
  Astrophys.\ J.\  {\bf 610}, L73 (2004).
  [astro-ph/0404137].

\bibitem{Hires2} Talk of S. Westerhoff at the CRIS-2004 workshop ``GZK and
Surrounding'', Catania, Italy,\\
\texttt{http://www.ct.infn.it/cris2004/talk/westerhoff.pdf}, also
in Nucl.\ Phys.\ Proc.\ Suppl.\  {\bf 136C} (2004) 46
  [astro-ph/0408343].





\bibitem{Sommers:2000us}
  P.~Sommers,
``Cosmic Ray Anisotropy Analysis with a Full-Sky Observatory,''
  Astropart.\ Phys.\  {\bf 14}, 271 (2001)
  [astro-ph/0004016].


\bibitem{Stanev:1995my}
  T.~Stanev, P.~L.~Biermann, J.~Lloyd-Evans, J.~P.~Rachen and A.~Watson,
  ``The Arrival directions of the most energetic cosmic rays,''
  Phys.\ Rev.\ Lett.\  {\bf 75} (1995) 3056
  [astro-ph/9505093];
See also the study by  G.~A.~Medina-Tanco,
  ``Large scale distribution of matter in the nearby universe and  ultra-high
  energy cosmic rays,'' astro-ph/9707054.


\bibitem{Dolag:2003ra}
K.~Dolag, D.~Grasso, V.~Springel and I.~Tkachev,
 ``Mapping deflections of Ultra-High Energy Cosmic Rays in Constrained
Simulations of Extragalactic Magnetic Fields,'' JETP Lett.\  {\bf
79}, 583 (2004) [Pisma Zh.\ Eksp.\ Teor.\ Fiz.\ {\bf 79}, 719
(2004)] [astro-ph/0310902];
See also
 ``Constrained simulations of the magnetic field in the
 local universe and the propagation of UHECRs,''
 astro-ph/0410419.

\bibitem{ems}
G.~Sigl, F.~Miniati and T.~En{\ss}lin,
 ``Ultra-High Energy Cosmic
 Ray Probes of Large Scale Structure and Magnetic Fields''
 Phys.\ Rev.\ D {\bf 70}, 043007 (2004) [astro-ph/0401084];
see also ``Cosmic magnetic fields and their influence on
ultra-high energy
 cosmic ray propagation,''
  Nucl.\ Phys.\ Proc.\ Suppl.\  {\bf 136} (2004) 224,
  astro-ph/0409098.







\bibitem{Kachelriess:2005qm}
  M.~Kachelriess, P.~D.~Serpico and M.~Teshima,
``The galactic magnetic field as spectrograph for ultra-high energy
cosmic rays,''
  Astropart.\ Phys.\  {\bf 26}, 378 (2006)
  [astro-ph/0510444].

\bibitem{Takami:2005ij}
  H.~Takami, H.~Yoshiguchi and K.~Sato,
  ``Propagation of ultra-high energy cosmic rays above 10**19-eV in a
  structured extragalactic magnetic field and galactic magnetic field,''
  Astrophys.\ J.\  {\bf 639}, 803 (2006)
  [Erratum-ibid.\  {\bf 653}, 1584 (2006)]
  [astro-ph/0506203].

\bibitem{Harari:1999it}
  D.~Harari, S.~Mollerach and E.~Roulet,
  ``The toes of the ultra high energy cosmic ray spectrum,''
  JHEP {\bf 9908} (1999) 022,
  astro-ph/9906309.



\bibitem{Harari:2006uy}
  D.~Harari, S.~Mollerach and E.~Roulet,
``On the ultra-high energy cosmic ray horizon,''
  JCAP {\bf 0611} (2006) 012
  [astro-ph/0609294].


\bibitem{Hooper:2006tn}
  D.~Hooper, S.~Sarkar and A.~M.~Taylor,
  ``The intergalactic propagation of ultra-high energy cosmic ray nuclei,''
  Astropart.\ Phys.\  {\bf 27} (2007) 199
  [astro-ph/0608085].




\bibitem{Sigl:1998dd}
  G.~Sigl, M.~Lemoine and P.~Biermann,
  ``Ultra-high energy cosmic ray propagation in the local supercluster,''
  Astropart.\ Phys.\  {\bf 10}, 141 (1999)
  [astro-ph/9806283].

\bibitem{Lemoine:1999ys}
  M.~Lemoine, G.~Sigl and P.~Biermann,
``Supercluster Magnetic Fields and Anisotropy of Cosmic Rays above
10$^{19}$ eV,''
  arXiv:astro-ph/9903124.


\bibitem{Deligny:2003rp}
  O.~Deligny, A.~Letessier-Selvon and E.~Parizot,
  ``Magnetic Horizons Of Uhecr Sources And The Gzk Feature,''
  Astropart.\ Phys.\  {\bf 21} (2004) 609
  [astro-ph/0303624].


\bibitem{Parizot:2004wh}
  E.~Parizot,
  ``GZK horizon and magnetic fields,''
  Nucl.\ Phys.\ Proc.\ Suppl.\  {\bf 136} (2004) 169
  [astro-ph/0409191].


\bibitem{Armengaud:2004yt}
  E.~Armengaud, G.~Sigl and F.~Miniati,
``Ultrahigh energy nuclei propagation in a structured, magnetized
universe,''
  Phys.\ Rev.\  D {\bf 72}, 043009 (2005)
  [astro-ph/0412525].


\bibitem{Burgett:2003yg}
  W.~S.~Burgett and M.~R.~O'Malley,
  ``Hints of energy dependences in AGASA EHECR arrival directions,''
  Phys.\ Rev.\  D {\bf 67} (2003) 092002
  [arXiv:hep-ph/0301001].


\bibitem{Cuoco:2007id}
  A.~Cuoco, S.~Hannestad, T.~Haugboelle, M.~Kachelriess and P.~D.~Serpico,
  ``Clustering properties of ultrahigh energy cosmic rays and the search for
  their astrophysical sources,''
  arXiv:0709.2712 [astro-ph].



\bibitem{Unger:2007mc}
  M.~Unger [Pierre Auger Collaboration],
``Study of the Cosmic Ray Composition above 0.4 EeV using the
Longitudinal Profiles of Showers observed at the Pierre Auger
Observatory,''
  arXiv:0706.1495 [astro-ph].

\bibitem{Gorski:2004by}
  K.~M.~Gorski, E.~Hivon, A.~J.~Banday, B.~D.~Wandelt, F.~K.~Hansen, M.~Reinecke and M.~Bartelman,
   ``HEALPix -- a Framework for High Resolution Discretization, and Fast
  Analysis of Data Distributed on the Sphere,''
  Astrophys.\ J.\  {\bf 622} (2005) 759
  [astro-ph/0409513].



\end{thebibliography}
\end{document}